\documentclass[sn-aps]{sn-jnl}

\usepackage{graphicx}%
\usepackage{multirow}%
\usepackage{amsmath,amssymb,amsfonts}%
\usepackage{amsthm}%
\usepackage{mathrsfs}%
\usepackage[title]{appendix}%
\usepackage{xcolor}%
\usepackage{textcomp}%
\usepackage{manyfoot}%
\usepackage{booktabs}%
\usepackage{algorithm}%
\usepackage{algorithmicx}%
\usepackage{algpseudocode}%
\usepackage{listings}%

\usepackage{amsmath} 

\usepackage{ae}
\usepackage{siunitx}
\usepackage{tikz}
\usetikzlibrary{quantikz}
\usepackage{subcaption}
\usepackage{braket}
\usepackage{amsthm}
\usepackage{amssymb}

\usepackage{comment}

\theoremstyle{thmstyleone}%
\theoremstyle{thmstyletwo}%

\theoremstyle{thmstylethree}%

\begin{document}

\title{Challenges and opportunities in the supervised learning of quantum circuit outputs}

\author*[1,2]{\fnm{Simone} \sur{Cantori}}\email{simone.cantori@unicam.it}
\author[1,2]{\fnm{Sebastiano} \sur{Pilati}}\email{sebastiano.pilati@unicam.it}

\affil[1]{\orgdiv{School of Science and Technology, Physics Division}, \orgname{Universit{\`a}  di Camerino}, \orgaddress{\street{Via Madonna delle Carceri}, \city{Camerino (MC)}, \postcode{62032}, \country{Italy}}}

\affil[2]{\orgdiv{Sezione di Perugia}, \orgname{INFN},  \city{Perugia}, \postcode{06123}, \country{Italy}}

\abstract{
Recently, deep neural networks have proven capable of predicting some output properties of relevant random quantum circuits, indicating a strategy to emulate quantum computers alternative to direct simulation methods such as, e.g., tensor-network methods. However, the reach of this alternative strategy is not yet clear.
Here we investigate if and to what extent neural networks can learn to predict the output expectation values of circuits often employed in variational quantum algorithms, namely, circuits formed by layers of CNOT gates alternated with random single-qubit rotations.
On the one hand, we find that the computational cost of supervised learning scales exponentially with the inter-layer variance of the random angles. This allows entering a regime where quantum computers can easily outperform classical neural networks. On the other hand, circuits featuring only inter-qubit angle variations are easily emulated. In fact, thanks to a suitable scalable design, neural networks accurately predict the output of larger and deeper circuits than those used for training, even reaching circuit sizes which turn out to be intractable for the most common simulation libraries, considering both state-vector and tensor-network algorithms.
We provide a repository of testing data in this regime, to be used for future benchmarking of quantum devices and novel classical algorithms.
}


\maketitle

\section{Introduction}
%
In recent years, deep-learning techniques have proven fitted to tackle various critical computational tasks in quantum many-body physics, including, among others~\cite{RevModPhys.91.045002,doi:10.1080/23746149.2020.1797528, schutt2020machine,Kulik_2022}, approximating ground-state wave functions~\cite{doi:10.1126/science.aag2302,PRXQuantum.2.040201}, learning universal energy-density functionals~\cite{PhysRevLett.108.253002,https://doi.org/10.1002/qua.25040,Brockherde2017,PhysRevE.106.045309}, implementing force fields for molecular dynamics~\cite{PhysRevLett.98.146401,10.1063/1.4966192}, or performing quantum-state tomography~\cite{torlai2018neural}.
It is therefore natural to ask whether neural networks can also emulate the behaviour of quantum computers. For example, they could learn to reproduce the complex correlations among qubits from sets of measured bitstrings~\cite{melko2024language}, or they could be trained via supervised learning to predict output expectation values from circuit descriptors~\cite{Cantori_2023}.

In many applications of supervised deep-learning to many-body problems, scalability has emerged as a critical enabling feature of the adopted network architectures~\cite{C8SC04578J,saraceni,https://doi.org/10.1002/syst.201900052,10.21468/SciPostPhys.10.3.073,Cantori_2023,PhysRevB.108.125113,blania2022deep,PhysRevE.108.065304}.
In the context of quantum computing, scalability might allow the trained networks to emulate larger circuits than those used for training, potentially reaching intractable regimes for other classical simulation methods. 
If successful, this approach could provide useful benchmark data for the further development of quantum computing devices.
In fact, some encouraging results have been reported in Ref.~\cite{Cantori_2023}. Specifically, the supervised training of deep networks was performed on small universal circuits amenable to classical simulations. The networks learned to map suitable representations of universal quantum circuits to the corresponding output expectation values, and they could be tested also on larger circuits.
Yet, the actual potential of this supervised-learning approach is not clear.  A performance degradation was in fact reported, e.g., for deeper circuits rather than for more qubits, or for scrambling rather than for localized circuit dynamics~\cite{mohseni2023}.
It is worth pointing out that assessing the potential of the supervised learning of quantum circuits is important also for the further development of quantum error mitigation schemes~\cite{Baireuther2018machinelearning,Chamberland_2018,Baireuther_2019,zlokapa2020deep,8880492}. 
For example, in Ref.~\cite{ml_qem}, error suppression was implemented by training machine-learning models on exact expectation values computed for classically tractable circuits. Chiefly, for intractable circuit sizes, training was performed on noisy data improved via conventional zero-noise extrapolation, leading to a reduction of the computational overhead, in particular at runtime.

The central goal of this article is to delineate some of the successes and the limitations of scalable supervised learning of the output properties of quantum circuits. We address this issue focusing on a general circuit structure familiar from many applications of the variational quantum eigensolver (see, e.g.,~\cite{doi:10.1021/acs.jctc.1c00091,PhysRevA.102.062612,Barkoutsos2020improving,shen2023prepare,scriva2023challenges}). Specifically, these circuits feature layers of CNOT gates alternated with single-qubit rotations with random angles.
On the one hand, we demonstrate a dramatic failure of classical supervised learning: the computational cost exponentially increases with the variance of the inter-layer angle fluctuations. This observation allows us pinpointing a regime where quantum devices can conveniently outperform classical deep learning. On the other hand, for circuits featuring only intra-layer angle variations, supervised training is efficient in terms of required training circuits. Moreover, the scalable networks accurately extrapolate to larger and deeper circuits than those used in the training phase. 
An analysis of the computational cost of state-vector and tensor-network simulations is provided, considering two of the most popular and performant software executed on high-performance computers. This analysis shows that the scalable networks largely overcome the range where the current general-purpose implementations of these classical algorithms are practical. To favour future developments of quantum devices and novel classical algorithms, a dataset of benchmark data produced by one of our trained neural networks is provided at the repository of Ref.~\cite{cantori_2024_10610695}.

The rest of the Article is organized as follows:
in Section~\ref{Sec2}, we describe the quantum circuits we consider, as well as the adopted neural-network architectures and their training protocols.
Section~\ref{Sec3} is devoted to the analysis of the prediction accuracy after supervised training, both for fixed circuit sizes and for extrapolations to larger and deeper circuits.
In Section~\ref{Sec4}, we discuss the computational cost of classical simulations of the quantum circuits, considering both state-vector simulations and a tensor-network method.
Section~\ref{Sec5} summarizes our main findings and their implications  on the perspective of simulating quantum computers via deep learning.

\begin{figure*}[]
	\centering
	\includegraphics[width=\textwidth]{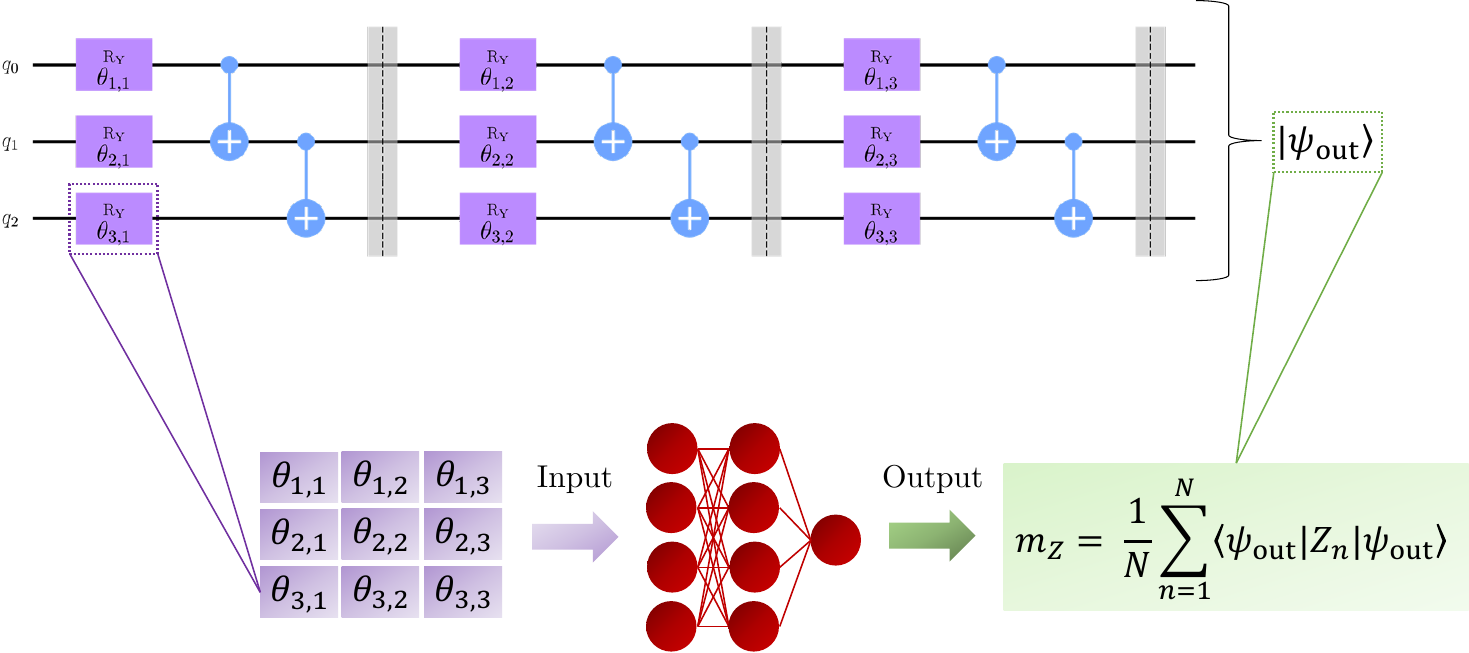}
	\caption{Schematic representation of the supervised learning protocol of quantum circuits.
 The structure of the quantum circuits is shown in the upper part. Single-qubit rotations $R_y(\theta_{n,p})$ with random angles $\theta_{n,p}$ at qubit $n=1,\dots,N$ and layer $p=1,\dots,P$ are alternated with CNOT gates acting on adjacent qubits. 
 The input state is $\left|\psi_{\mathrm{in}}\right>=\left|0\right>^{\otimes N}$.
 The circuit descriptors $\theta_{n,p}$ (shown in the lower part) constitute the neural network input. The output is the expectation value $m_z = \frac{1}{N}\sum_{n=1}^{N}\left<\psi_{\mathrm{out}}\right| Z_n \left|\psi_{\mathrm{out}}\right>$, where $\left|\psi_{\mathrm{out}}\right>$ is the output state of the quantum circuit.}
	\label{Fig1}
\end{figure*}

\section{Methods}\label{Sec2}
\subsection{Quantum circuit architecture}
\label{secQC}
\begin{figure*}[]
	\centering
	\includegraphics[width=\textwidth]{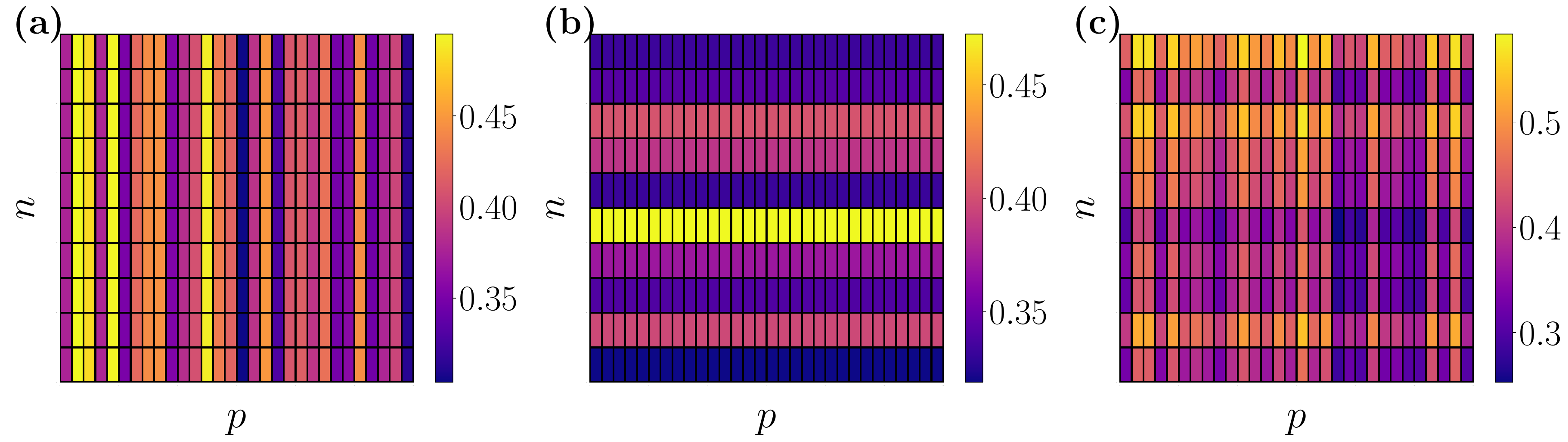}
	\caption{Color-scale representation of the random angle $\theta_{n,p}$ as a function of the qubit index $n=1,\dots,N$ and the layer index $p=1,\dots,P$. The angles are sampled according to Eq.~\eqref{eqangles}.
 Three circuit configurations are considered:  
 i)  Only inter-layer variations are allowed by setting $\sigma^2_N=0$, see panel (a). 
 ii) Only intra-layer variations are allowed by setting  $\sigma^2_P=0$, see panel (b). 
 iii) Both intra-layer and inter-layer variations are allowed, see panel (c).
 }
	\label{QC}
\end{figure*}

We consider quantum circuits composed of $N$ qubits and $P$ layers of gates. 
In each layer, a parametrized single-qubit gate is applied to each qubit, and the two-qubit CNOT gate is applied to all pairs of adjacent qubits (assuming open boundary conditions). 
The single-qubit gates implement rotations around the $y$ axis, corresponding to the following matrix representation:
\begin{equation}\label{Ry}
    R_y(\theta)=\begin{bmatrix}
			\cos{\frac{\theta}{2}} & -\sin{\frac{\theta}{2}} \\
			\sin{\frac{\theta}{2}} & \cos{\frac{\theta}{2}} 
		\end{bmatrix}\, .
\end{equation}
Notice that the angles applied to different qubits or in different layers may vary; we denote with $\theta_{n,p}$ the angle for the qubit with index $n=1,\dots,N$ at the layer $p=1,\dots,P$.
These angles are randomly generated as follows:
\begin{equation}
\label{eqangles}
    \theta_{n,p} = u_n + u_p,
\end{equation}
where $u_{n/p}\sim U\left(a_{N/P},b_{N/P}\right)$, and the symbol $U\left(a,b\right)$ denotes the uniform distribution in the range $u\in \left[a,b\right]$. In the following, we consider intra-layer and inter-layer (uniform) distributions with means $\mu_N$ and $\mu_P$, respectively;  the corresponding variances are denoted as $\sigma_N^2$ and $\sigma_P^2$.
The required boundaries of the uniform ranges are easily determined as $a_{N/P} = \mu_{N/P}-\sqrt{3\sigma_{N/P}^2}$ and $b_{N/P} = \mu_{N/P}+\sqrt{3\sigma_{N/P}^2}$.
Specifically, three circuit configurations are considered: 
i) only inter-later angle fluctuations are allowed, i.e., we set $\sigma_N=0$; 
ii) only intra-layer fluctuations are allowed, i.e., we set $\sigma_P=0$; 
iii) both intra-layer and inter-layer fluctuations are allowed, i.e., $\sigma_N>0$ and $\sigma_P>0$.
Examples of these three configurations are represented in Fig.~\ref{QC}.

To perform supervised learning, a suitable set of circuit descriptors, which we collectively indicate with the symbol $\Theta$, is needed~\cite{Zhang_2021}. For what concerns configuration i), the angles only depend on the layer index $p$, i.e., we can write $\theta_{n,p}=\theta_{p}$; therefore, each circuit is unambiguously identified by the $P$-dimensional array $\Theta=(\theta_{1},\dots,\theta_{P})$. For configuration ii), one can write $\theta_{n,p}=\theta_{n}$, and the minimal unambiguous circuit representation is the $N$-dimensional array $\Theta=(\theta_{1},\dots,\theta_{N})$. 
Within configuration iii), a $N\times P$ matrix is needed, namely we set $\Theta=\left[\theta_{n,p}\right]$.
Here, it is worth pointing out that the $N\times P$ descriptor matrix can, in principle, be adopted also for configurations i) and ii), by considering constant angles for different qubits or different layers, respectively. In fact, we employ this matrix representation in some of the extrapolation tests, performed on circuits in configurations ii), in subsection~\ref{subsec3b}. In this case, the $N\times P$ descriptors matrix reads:
\begin{equation}
\Theta= 
 \begin{bmatrix}
\theta_1 & \dots & \theta_1 \\
\theta_2 & \dots & \theta_2 \\
\vdots    & \ddots & \vdots    \\
\theta_N & \dots & \theta_N \\
\end{bmatrix}.
\end{equation}

\subsection{Neural network architecture}
\label{subsecNN}
Our goal is to train deep neural networks to learn the map $f_{\omega}\left(\Theta\right)=\tilde{y}$, where with $\omega$ we denote the weights and biases characterizing the network, and the output $\tilde{y}$ is supposed to accurately approximate the chosen target value.
Specifically, the target we consider is the average magnetization per qubit: 
\begin{equation}
    m_z = \frac{1}{N}\sum_{n=1}^{N}\left<\psi_{\textrm{out}}\right| Z_n \left| \psi_{\textrm{out}} \right> \, ,
\end{equation}
where $Z$ is the standard Pauli operator, and the output state can be computed as $\left|\psi_{\textrm{out}}\right>=\hat{U} \left|\psi_{\textrm{in}}\right>$, with $\hat{U}$ the unitary operator corresponding to the quantum circuit; the input state $\left|\psi_{\textrm{in}}\right>=\left|0\right>^{\otimes N}$ is a product state of spin-up eigenstates of $Z_n$.

Three neural-network architectures are considered: multilayer perceptrons (MLPs), convolutional neural networks (CNNs), and long short-term memory networks (LSTMs).
Hereafter, these models are briefly described, while all relevant further details are provided in Appendix~\ref{appendix}.
MLPs represent perhaps the most archetypal neural-network structure (they are sometimes referred to as ``plain vanilla'' models~\cite{hastie2009elements}). They are formed by a sequence of dense layers, i.e. layers featuring (bipartite) all-to-all connectivity among neurons in adjacent layers. The numbers of layers and of neurons in each layer can be adjusted, seeking for a compromise between expressivity and computational cost. This architecture is not scalable, i.e., it can be applied only to inputs of a predetermined fixed size. We adopt it only for the analysis of subsection~\ref{subsec3a}, where the size and depth of the testing circuits coincides with the ones of the training circuits.

As a term of comparison, we confront the performance of MLPs against the ones of more sophisticated models, namely, the CNNs and LSTMs. 
The latter represent a type of recurrent neural network. They are commonly adopted in the analysis of temporal sequences. The long-short memory structure is designed to account for memory effect, while suppressing the vanishing gradient problems~\cite{10.1162/neco.1997.9.8.1735}. Their key feature is the LSTM cell, which includes input, forget, and output gates. These gates control the flow of information, enabling the network to selectively store, discard, and utilize information over extended sequences.
Recently, LSTMs have been adopted also to learn the unitary dynamics of quantum spin models, adiabatic quantum computers, and gate-based circuits~\cite{lstm,global_pooling2,mohseni2023}.
We also apply them to quantum circuits, specifically for configuration i), by identifying the (discretized) time with the layer index $p$. The angle $\theta_p$ is used to predict the expectation value at the corresponding time index $m_z^{(p)}$, together with the information passed from the previous time history. The resulting scheme is detailed in Appendix~\ref{appendix}.

In subsection~\ref{subsec3b}, we perform extrapolation tests, namely, we test the trained networks on larger and deeper circuits than those included in the training sets. For this, a scalable architecture is required, pointing to convolutional models. Yet, it is worth pointing out that the standard CNNs adopted in computer science (usually for image recognition~\cite{venkatesan2017convolutional}), which feature convolutional filters with fixed-size kernels followed by a set of dense layers, are not fully scalable. In fact, while the former layers can, in principle, be applied to variable-size inputs, dense layers require instead a fixed input size. Following Ref.~\cite{saraceni}, we include a global pooling operation after the last convolutional layer, thus obtaining a fully scalable architecture. 
We perform extrapolations, with circuits in configuration ii), to larger qubit numbers at fixed circuit depths, as well as extrapolations to both larger and deeper circuit. In the first case, a one-dimensional CNN operating on the $N$-dimensional descriptor array is appropriate. In the second case, scalability in both $N$ and $P$ is obtained considering a two-dimensional CNN, and the input is a $N \times P$ matrix, as discussed in subsection~\ref{secQC}.
It should be mentioned here that, in the literature, various alternative strategies have been discussed to perform scalable learning of quantum many-body systems~\cite{10.21468/SciPostPhys.10.3.073}. These strategies include, e.g., 
tiling the systems~\cite{C8SC04578J}, adopting fixed-size system representations~\cite{https://doi.org/10.1002/syst.201900052}, or addressing extensive observables~\cite{blania2022deep,lstm}.
For the tests on fixed circuit sizes and depths, full scalability is instead not required. Hence, a standard CNN featuring a flatten operation to connect the last convolutional layer to the dense part is adequate. We adopt this structure for the tests in subsection~\ref{subsec3a}.

The training of the networks described above is performed via supervised learning. 
A training dataset 
$\{(\Theta_k,y_k)\}_{k=1}^{K_\mathrm{train}}$, 
where $y_k$ represents the exact target value corresponding to the circuit with angles $\Theta_k$, is generated by performing circuit simulations using the Qiskit library~\cite{qiskit} and the qsimcirq library~\cite{qsimcirq}. 
We mostly perform exact state-vector simulations. Tensor-network methods are also adopted to assess the cost of classical simulations; see Section~\ref{Sec5}.
The network parameters $\omega$ are optimized by minimizing the mean squared error
\begin{equation}\label{mse}
	\mathcal{L} = \frac{1}{K_\mathrm{train}}\sum_{k=1}^{K_\mathrm{train}}
 \left(y_k - \tilde{y}_k \right)^2 \, ,
\end{equation}
where $\tilde{y}_k=f_{\omega}(\Theta_k)$. 
The training is performed via a popular variant of stochastic gradient descent, namely, the ADAM algorithm~\cite{adam}.
To quantify the prediction accuracy, we consider the coefficient of determination
\begin{equation}\label{r2}
R^2=
1-\frac{
\sum_{k=1}^{K_\mathrm{test}} \left(y_k - \tilde{y}_k \right)^2}{\sum_{k=1}^{K_\mathrm{test}} \left(y_k - \bar{y} \right)^2} ,
\end{equation}
where $\bar{y}$ is the average of the target values and $K_{\mathrm{test}}$ is the number of random circuits in the test set. Clearly, the test sets contain only circuits which are not included in the training set. Notice that perfect predictions correspond to $R^2=1$, while a constant model with the exact average corresponds to $R^2=0$.

\section{Accuracy of supervised learning}\label{Sec3}
\subsection{Testing on fixed circuit sizes}
\label{subsec3a}
\begin{figure}[]
	\centering
	\includegraphics[width=\columnwidth]{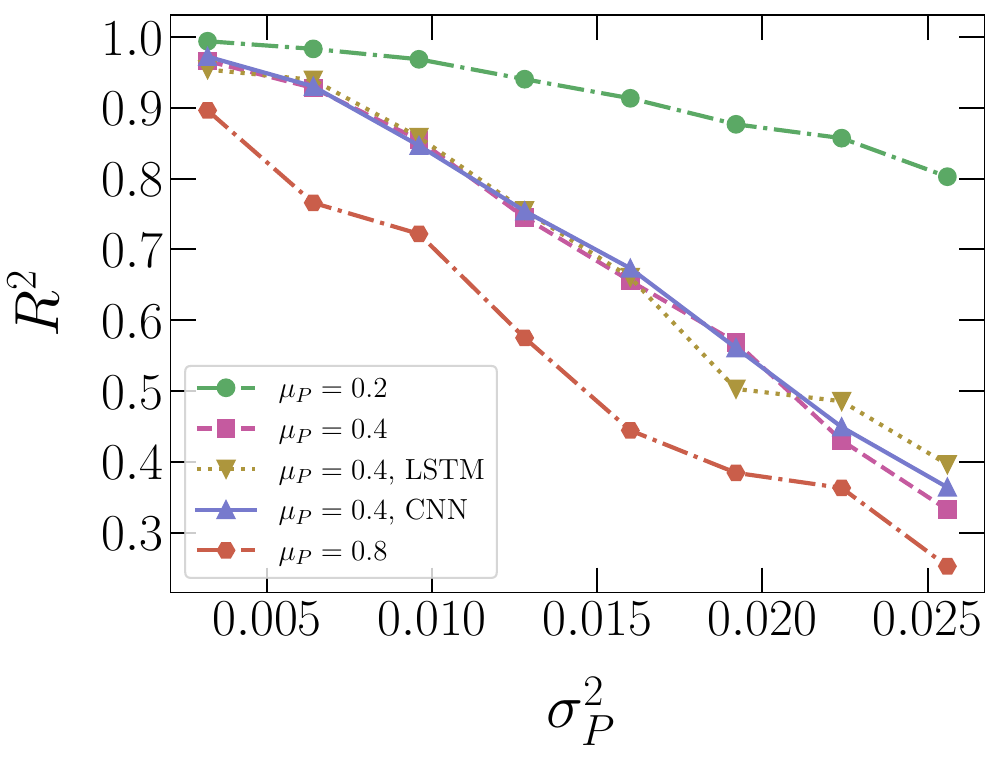}
	\caption{Coefficient of determination $R^2$ as a function of the inter-layer variance $\sigma_P^2$. Different datasets correspond to different values of the mean $\mu_P$ and to different network architectures: MLP, LSTM, and CNN. 
    The networks are trained and tested on quantum circuits of the same size:  $N=10$ and $P=30$. The training sets include $10^4$ circuits.}
	\label{TestbedA_QC_A}
\end{figure}

Hereafter we investigate if and in which conditions deep learning techniques allow predicting the magnetization per spin $m_z$ in the output state of the quantum circuits described in subsection~\ref{secQC}. In this first analysis, the tests are performed on circuits with the same qubit number and circuit depth of the ones included in the training datasets.
Hence, a scalable architecture is not required.
Most of the tests are performed by adopting the MLP architecture, but LSTMs and CNNs are also considered for comparison.
The first test addresses circuits in configuration i) (see subsection~\ref{secQC}), which only feature inter-layer angle fluctuations. The results are shown in Fig.~\ref{TestbedA_QC_A}.
One notices that the prediction accuracy rapidly decreases with the inter-layer variance $\sigma_P^2$. The drop is somehow delayed for smaller angle means $\mu_P$~\footnote{It occurs also for $\mu_P\simeq 0$, but in a larger $\sigma_P$ range than the one shown in Fig.~\ref{TestbedA_QC_A}.}.
Furthermore, the three adopted neural networks (MLP, CNN, and LSTM) display very similar performances, indicating that the network architecture does not play a key role.

\begin{figure}[]
	\centering
	\includegraphics[width=\columnwidth]{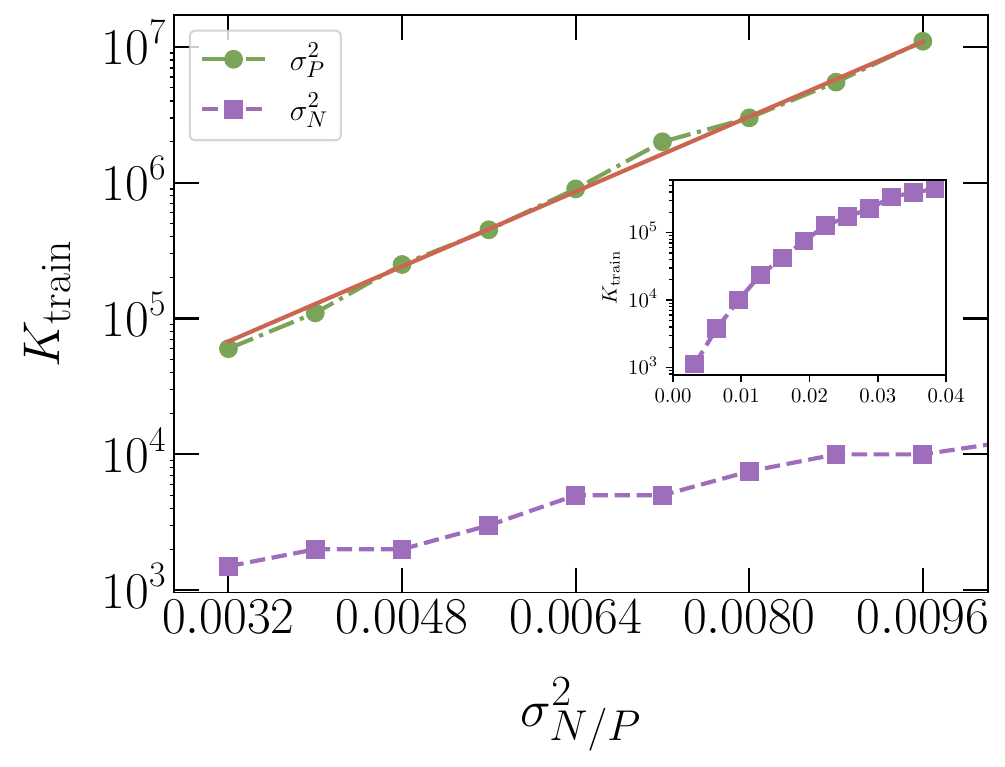}
	\caption{Number of training circuits $K_{\mathrm{train}}$ required to reach the accuracy $R^2\ge 0.99$, as a function of the inter-layer variance $\sigma^2_P$ or the intra-layer variance $\sigma^2_N$ . 
 The MLP is trained and tested on circuits with $N=10$ qubits and $P=30$ layers. 
 (Green) circles correspond to circuits in configuration i):  $\sigma_N^2=0$ and inter-layer mean $\mu_P=0.4$; (purple) squares correspond to configuration ii): $\sigma_P^2=0$ and $\mu_N=0.4$. The continuous (red) line represents an exponential fitting function: $K_{\mathrm{train}}=a\exp(b\sigma^2_P)$, with the parameters $a\simeq5350$ and $b\simeq794$ obtained from a best fit analysis. The inset displays the scaling of $K_{\mathrm{train}}$ in a larger range of $\sigma^2_N$; here, the error-bar is estimated considering two trainings of the MLP and it turns out to be smaller than the marker size.}
	\label{TestbedA_NT}
\end{figure}

In general, one expects that networks trained on larger datasets display a superior performance. This indeed occurs also in the problem addressed here. To quantify this effect, we determine the number of training circuits $K_{\textrm{train}}$ required to reach the accuracy threshold $R^2 \ge 0.99$, as a function of the inter-layer variance $\sigma_P^2$.
The results are analysed in Fig.~\ref{TestbedA_NT}, considering the MLP architecture. Notably, we observe an exponential increase as a function of $\sigma_P^2$, which in turn implies an exponential computational cost for dataset generation and network training. Already for $\sigma_P^2 \gtrsim 0.01$, computationally prohibitive numbers of training circuits are required. This indicates the failure of supervised learning for circuits with large inter-layer angle fluctuations.
Notice that this type of failure differs from the one discussed in Ref.~\cite{mohseni2023}, as the latter only occurs in extrapolations to larger circuit sizes, while here it occurs already for fixed circuit size.
%
%
%

\begin{figure}[]
	\centering
	\includegraphics[width=\columnwidth]{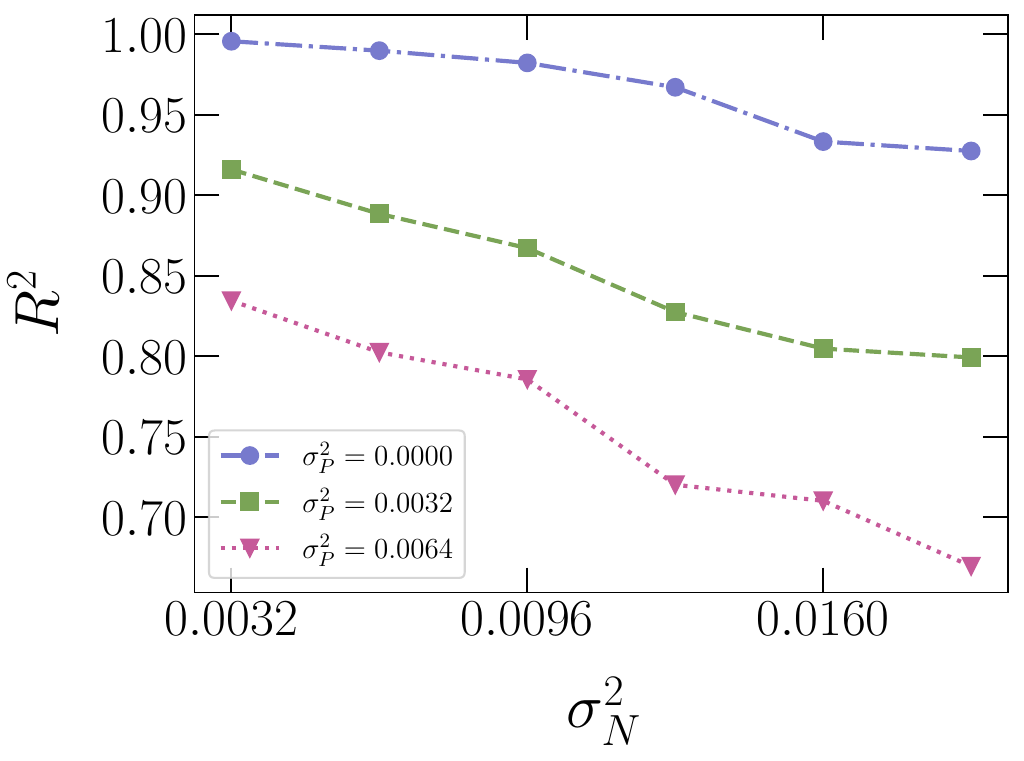}
	\caption{Coefficient of determination $R^2$ as a function of $\sigma_N^2$ for different values of $\sigma^2_P$. The MLPs are trained and tested on quantum circuits with $N=10$, $P=30$, $\mu_N=0.4$ and $\mu_P=0$. Here, the MLP has input shape $N\times P$. The training sets include $10^4$ quantum circuits.}
	\label{TestbedA_QC_A_varN}
\end{figure}

On the other hand, for circuits with small inter-layer variance $\sigma_P\simeq 0$, the network predictions are very accurate. 
This is confirmed by the analysis on circuits in configuration iii) (see details in Section~\ref{Sec2}), which feature both intra-layer and inter-layer angle fluctuations; this analysis is shown in Fig.~\ref{TestbedA_QC_A_varN}. One observes a slow accuracy decrease with the intra-layer variance $\sigma_N^2$, while increasing the inter-layer variance $\sigma_P^2$ causes a sharp downward shift.
Again, it is worth determining the required dataset size $K_{\mathrm{train}}$ to reach $R^2\ge0.99$, as a function of $\sigma_N^2$, now fixing $\sigma_P^2=0$. These results are shown in Fig.~\ref{TestbedA_NT}. One notices that orders of magnitude smaller $K_{\mathrm{train}}$ are required, and that the rate of increase is qualitatively slower as compared to the previously analysed case, i.e. configuration i). 
Hence, for relevant rotation ranges, e.g., in the range $\theta_{n,p}\in [0,\pi/2]$~\cite{ibm}, circuits featuring only intra-layer fluctuations are amenable to supervised learning. This is in contrast with the case of inter-layer fluctuations, where the required $K_{\mathrm{train}}$ can be easily tuned to unfeasible sizes.

\subsection{Extrapolations to larger circuits}
\label{subsec3b}
\begin{figure}[]
	\centering
	\includegraphics[width=\columnwidth]{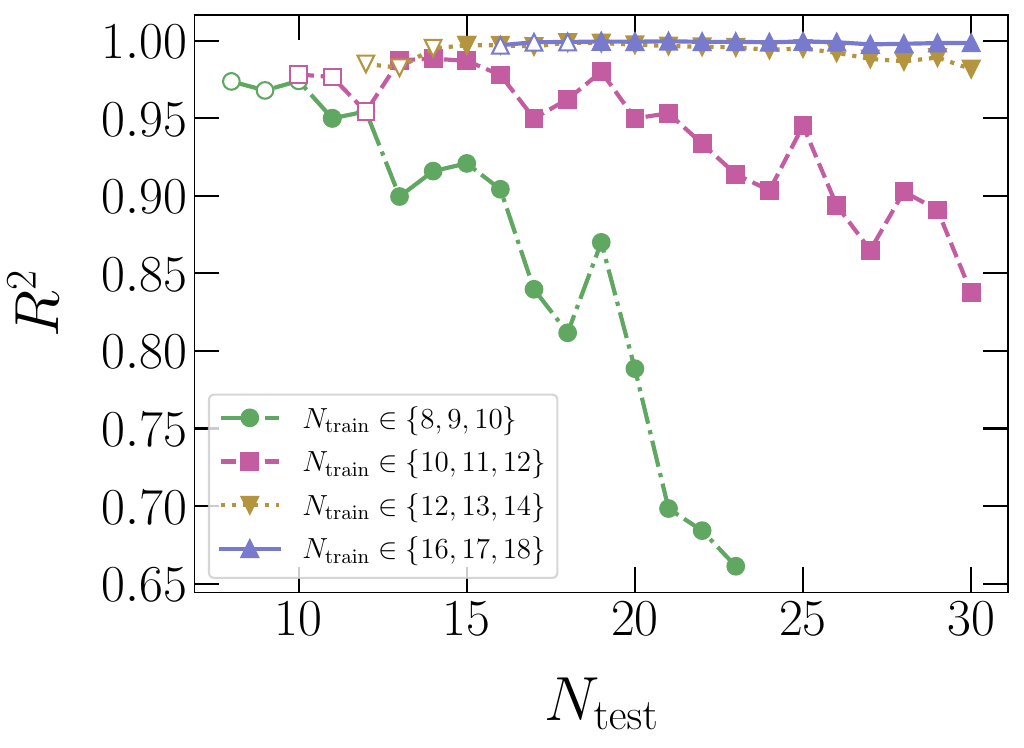}
	\caption{$R^2$ as a function of the number of qubits $N_{\mathrm{test}}$ of the testing quantum circuits. Four 1D-CNNs are trained on different qubit numbers $N_{\mathrm{train}}$, considering circuits in configuration ii) of depth $P=30$, with $\mu_N=0.4$ and $\sigma_N^2=0.1024$. The empty symbols represent the tests performed on the same circuit size as the training circuits. Filled symbols correspond to extrapolations to large circuits.}
	\label{TestbedB_N_extrap}
\end{figure}

Hereafter, we test the networks on larger and/or deeper circuits than those used in the training phase. For this, a scalable network architecture is required. We consider circuits in configuration ii), i.e., we allow only intra-layer angle variations by setting $\sigma_P=0$. In this scenario supervised learning for fixed circuit sizes has proven successful (see subsection~\ref{subsec3a}), and it is thus at least plausible to obtain accurate extrapolations also for larger and deeper circuits.
In the first test, the extrapolation is performed only to larger qubit numbers, fixing the depth of both training and testing circuits at $P=30$. 
In this setup, a one-dimensional CNN provides the required scalability, as discussed in subsection~\ref{subsecNN}.
The random angles are sampled from uniform distributions with mean $\mu_N=0.4$ and variance $\sigma^2_N=0.1024$. 
%
To help the networks learning how to scale the predictions with the circuit size, we gather heterogeneous training datasets featuring circuits with different qubit numbers. Specifically, the training qubit numbers are $N_{\mathrm{train}}\in\{N_0,N_0+1,N_0+2\}$, and the smallest training circuits range from $N_0=8$ to $N_0=18$ qubits. Here, the training sets include $1.2\times10^6$ circuits for each size. 
In Fig.~\ref{TestbedB_N_extrap}, the prediction accuracy is analysed as a function of the size of the testing circuits $N_{\mathrm{test}}$. 
It is reasonable to expect that small training circuits might not provide sufficient information to allow the network accurately extrapolating to qubit numbers $N_{\mathrm{test}} \gg  N_{\mathrm{train}}$. This is indeed the case; for example, the network trained with $N_0=8$ is no longer accurate in the regime $N_{\mathrm{test}} \gtrsim 13$. On the other hand,  when the training qubit numbers are sufficiently large, namely $N_{\mathrm{train}} \gtrsim 12$, the accuracy no longer collapses for larger  $N_{\mathrm{test}}$. This allows one to accurately predict the output of circuits with, at least, as many as $N_{\mathrm{test}}=30$ qubits. 
For the depths we consider, performing state-vector simulations of significantly larger circuits on classical computers is impractical. In fact, already to reach qubit numbers $N_{\mathrm{test}}\gtrsim 25$, high-performance computers are required. Furthermore, for the circuit model addressed here, simulations based on tensor-network methods are also ruled out, since they require impractically large bond-dimensions. 
This point is demonstrated in Section~\ref{Sec4}, where we also provide timings and other relevant details of our state-vector simulations in the large circuit regime.

\begin{figure}[]
	\centering
	\includegraphics[width=\columnwidth]{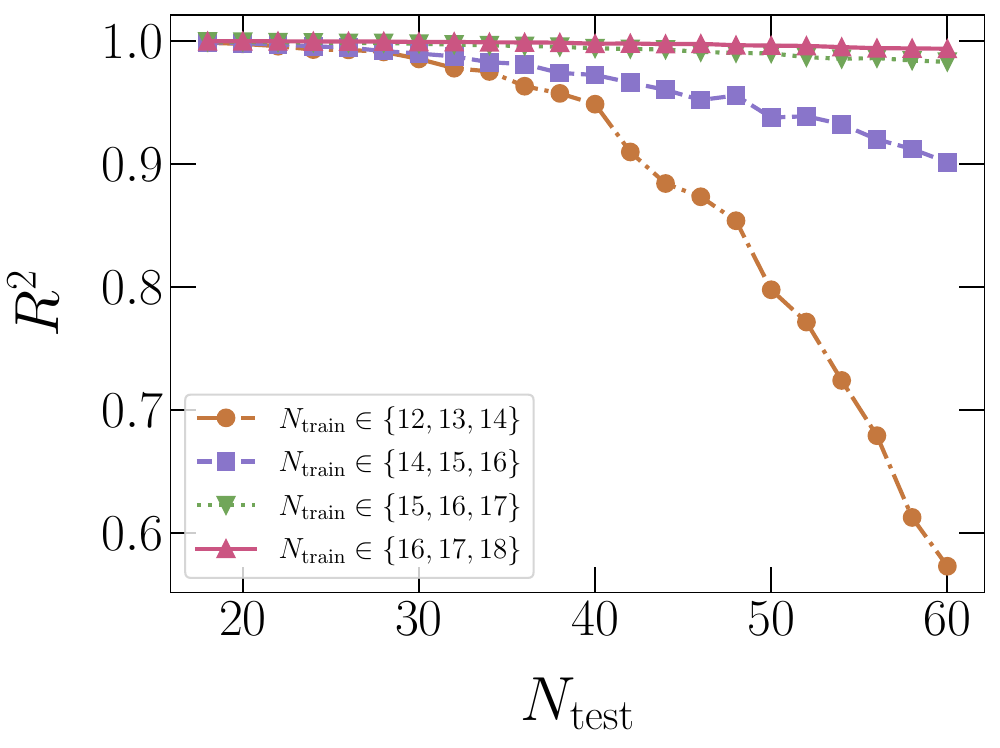}
	\caption{$R^2$ as a function of the number of qubits $N_{\mathrm{test}}$ of the testing circuits. Four 1D-CNN are trained on heterogeneous circuits featuring three qubit numbers $N_{\mathrm{train}}$ (see keys), and their predictions are compared to the ones from the training sizes $N_{\mathrm{train}}\in \{18,19,20\}$. The circuits are in configuration ii), with angles as in Fig.~\ref{TestbedB_N_extrap}.}
	\label{extrap_comparison}
\end{figure}

In the absence of benchmark data from other classical simulation methods in the regime $N_{\mathrm{test}} \gtrsim 30$, the neural-network accuracy can be assessed by comparing the predictions provided from trainings performed on different qubit numbers $N_{\mathrm{train}}$. 
In Fig.~\ref{extrap_comparison}, we examine the coefficient of determination $R^2$ of different networks, identifying the predictions of the network with the largest training circuits, namely, $N_{\mathrm{train}} \in \{18,19,20\}$, as the ground-truth benchmark. One notices that the predictions rapidly converge as the training circuit-size increases. The predictions obtained from the second-largest training set barely deviate from the chosen benchmark, reaching $R^2 > 0.99$ even up to $N_{\mathrm{test}}=60$. This accuracy threshold can be interpreted as a conservative estimate of the prediction accuracy of the benchmark network in the regime $N_{\mathrm{test}}\simeq 60$. 

Thus, we conclude that scalable supervised learning provides reliable expectation-value predictions in regime where the current implementations of two of the most popular classical simulation methods are essentially impractical.
To favour future comparisons with both quantum devices and novel classical simulation platforms, 
the predictions for quantum circuits in the ramge $N \in [30, 60]$ are made available to the community through the repository in Ref.~\cite{cantori_2024_10610695}.
It is worth emphasizing that this network is trained and tested on random angles with mean $\mu_N=0.4$ and (intra-layer) variance $\sigma^2_N=0.1024$. Tests performed on different distributions are not guaranteed to be successful. For example, performing the test on the circuit with no rotations, corresponding to the descriptors $\theta_n=0$ for $n=1,\dots,N$, the network prediction turns out to largely deviate from the expected result $m_z=1$. Yet, this problem can be easily solved, e.g., by augmenting the training dataset with $10^3$ circuits generated with mean $\mu_N=0$ and $\sigma_N^2=10^{-4}$. This simple trick leads to an accurate prediction of the CNN trained with quantum circuits with $N_\mathrm{train}\in\{16,17,18\}$ qubits for the above test circuit for $N_\mathrm{test}=100$, namely, $\tilde{y}=0.99074167 \cong m_z$.

\begin{figure}[]
	\centering
	\includegraphics[width=\columnwidth]{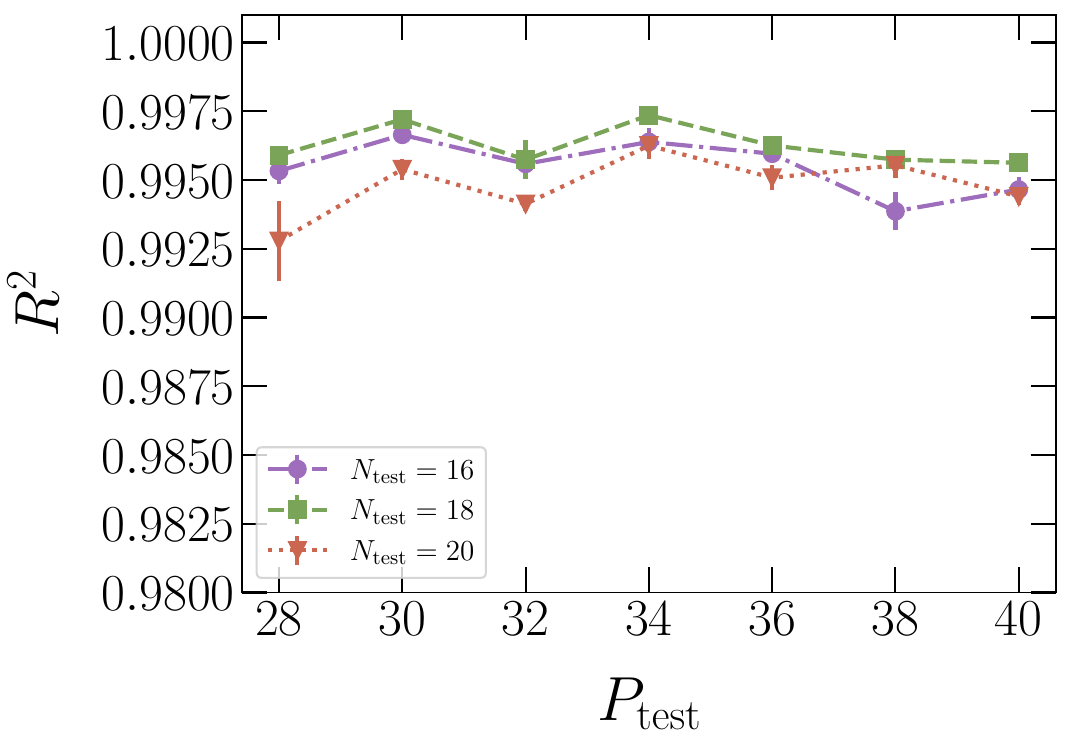}
	\caption{$R^2$ as a function of the number of layers of gates $P_{\mathrm{test}}$ of the quantum circuits in the test set, for different number of qubits $N_{\mathrm{test}}$. The CNN is trained on circuits in configuration ii) with $N_{\mathrm{train}} \in \{12,13,14\}$, $P_{\mathrm{train}} \in \{28,30\}$, $\mu_N=0.4$, and $\sigma_N^2=0.1024$. The test set includes $10^3$ circuits and it is divided in four subsets with $250$ instances. The error-bars represent the estimated standard deviation of the average over the four subsets.}
	\label{TestbedD_QC_B}
\end{figure}

\begin{figure}[]
	\centering
	\includegraphics[width=\columnwidth]{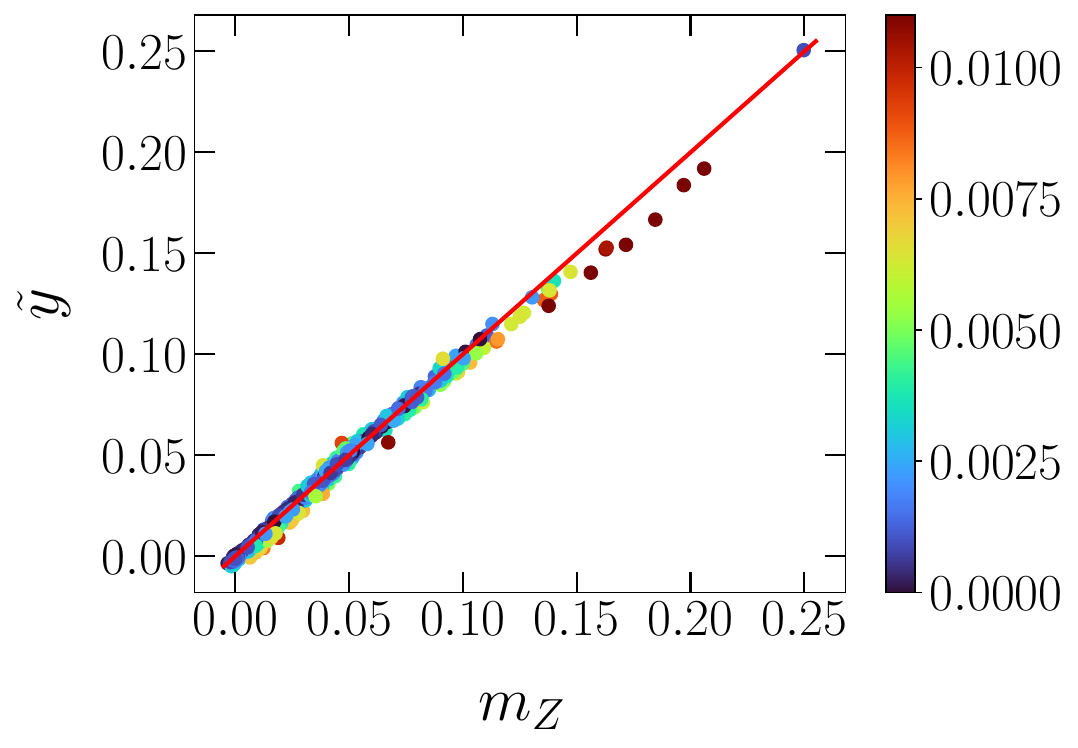}
	\caption{Magnetization per spin $\tilde{y}$ predicted by the CNN as a function of the exact results $m_z$. The quantum circuits have $N=20$ qubits and $P=40$ layers of gates. The CNN has been trained on quantum circuits with $N_{\mathrm{train}}\in\{12,13,14\}$ and $P_{\mathrm{train}}\in\{28,30\}$. The color scale represents $\left|\tilde{y}-m_z\right|$. The (red) line represents the bisector $\tilde{y} = m_z$. The rotation angles are as in Fig.~\ref{TestbedD_QC_B}.}
	\label{TestbedD_scatter}
\end{figure}

Next, we analyse extrapolations in two directions, namely, to circuits featuring both more qubits and more layers than those included in the training set. In this case, full scalability is obtained adopting a two-dimensional CNN, as discussed in subsection~\ref{subsecNN}. The extrapolation accuracy is analysed in Fig.~\ref{TestbedD_QC_B}. The network is trained on circuits with qubit numbers $N_{\mathrm{train}}\in \{12,13,14\}$ and depths $P_{\mathrm{train}}\in \{28, 30\}$, featuring $1.2\times10^6$ instances of each size and depth.
Notably, also in this test the prediction accuracy appears not to degrade with the size and the depth of the testing circuits, indicating that scalability in both directions is indeed feasible.
To further visualize the networks accuracy and the (non-trivial) output distribution, in Fig.~\ref{TestbedD_scatter} we show a scatter plot of predictions $\tilde{y}$ for circuits with $N_{\mathrm{test}}=20$ qubits and $P=40$ layers, as a function of the ground-truth expectation values $m_z$. The good agreement is clear. Some small deviations only occur for the largest magnetizations in the dataset. This can be attributed to the sparsity of training instances in that regime.


\section{Computational cost of classical circuit simulations}
\label{Sec4}
\begin{figure}[]
	\centering
	\includegraphics[width=\columnwidth]{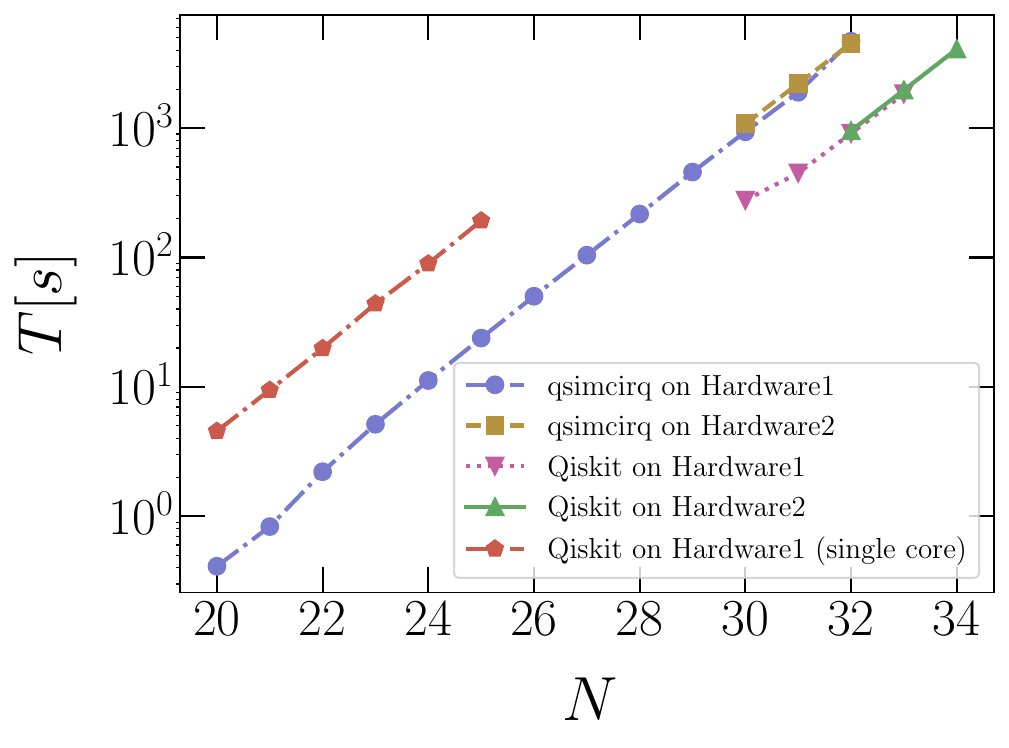}
	\caption{Time $T$ (in seconds) required to simulate one random circuit with state-vector methods, as a function of the qubit number $N$. The simulations are performed with the Qiskit and qsimcirq libraries on two different hardware. The circuits have $N$ qubits, $P=30$ layers of gates, $\mu_N=0.4$, $\sigma_N^2=0.1024$, and $\sigma_P^2=0$.}
	\label{statevector}
\end{figure}

The computational cost of accurate classical simulations of general quantum circuits is expected to scale exponentially for sufficiently large sizes and depths. Hereafter, we demonstrate that, for the random circuit we address, this exponential scaling sets in already for $N\gtrsim 20$, making simulations of $N\simeq 60$ circuits essentially impossible. 
First, we consider so-called state-vector methods, which provide numerically exact predictions of output properties. 
We adopt two of the most popular circuit simulation software, namely, the Qiskit~\cite{qiskit} and qsimcirq~\cite{qsimcirq} libraries.
It is worth mentioning that a detailed performance assessment of essentially all leading software platforms for circuit simulations has recently been reported~\cite{jamadagni2024benchmarking}. Considering variegate circuit types, that study assessed that Qiskit and qsimcirq are, indeed, the most efficient libraries in the large $N$ regime across variegate circuit types and hardware platforms. 
Notice that also Ref.~\cite{jamadagni2024benchmarking} presents exponential scalings for $N \gtrsim 20$, while timings on smaller circuits are often influenced by programming overheads.
In Fig.~\ref{statevector}, we show the  computation time for two different hardware, referred to as Hardware1 and Hardware2. Notice that qsimcirq operates using single precision, while Qiskit is set to operate in double precision. By fully utilizing Qiskit's capabilities, we are able to maximize the performance of the available hardware by exploiting all CPU cores. In particular, Hardware1 features the following CPU: Intel(R) Xeon(R) Gold 6154 CPU, 72 threads (36 cores with hyperthreading), with 192 Gb of RAM memory; Hardware2 is available from the Galileo100 cluster at CINECA, whereby each node features 2 Intel CascadeLake 8260 CPUs, with 24 cores each and a total of 384GB RAM memory.
Importantly, the largest qubit number we are able to address $N=34$ requires order of $T=5\times 10^3$s. Exploiting accelerators such as GPUs is expected to provide a quantitative speed-up~\cite{jamadagni2024benchmarking}, but sizes such as $N\simeq 60$ are evidently out of reach due to the observed exponential scaling with $N$.

\begin{figure}[]
	\centering
	\includegraphics[width=\columnwidth]{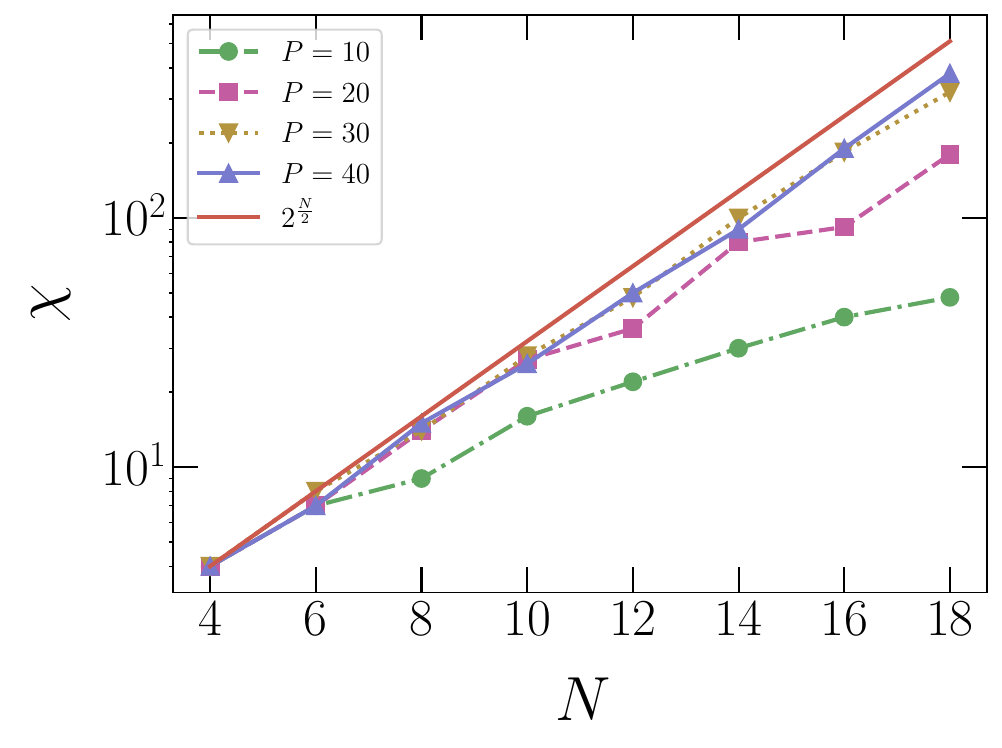}
	\caption{Bond dimension $\chi$ required to simulate quantum circuits with the accuracy $R^2\ge0.99$ using Qiskit's matrix product state (MPS) method, as a function of the qubit number $N$. To calculate $R^2$ we compare the MPS prediction with the ground-truth result $m_z$ obtained via a state-vector simulation. The results are averaged over 10 random circuits, featuring $P$ layers of gates, $\mu=0.4$, $\sigma_N^2=0.1024$, and $\sigma_P^2=0$. The (red) line represents the bond dimension needed to perform an ideal simulation: $\chi \propto 2^{\frac{N}{2}}$.}
	\label{MPS_var1024_mean4}
\end{figure}

\begin{figure}[]
	\centering
	\includegraphics[width=\columnwidth]{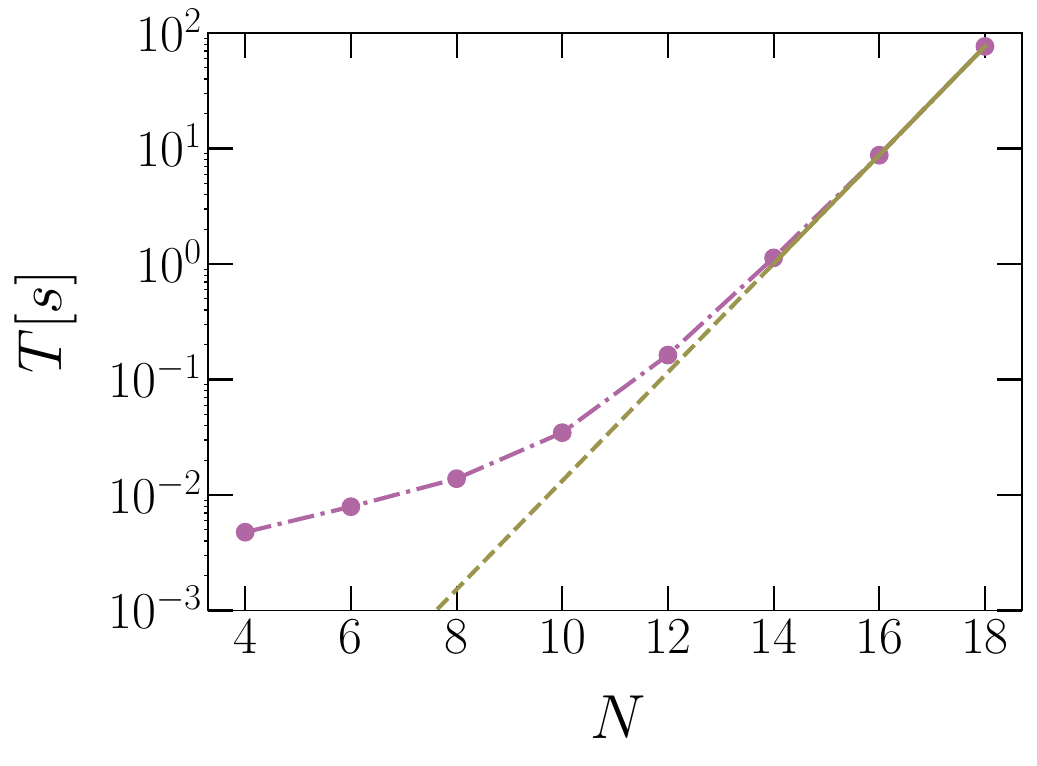}
	\caption{Time $T$ (in seconds) required to simulate, using Qiskit MPS method, one quantum circuit as a function of the number of qubits $N$. The circuits have $P=30$ layers of gates, $\mu_N=0.4$, $\sigma_N^2=0.1024$, and $\sigma_P^2=0$. The continuous (dark yellow) line represents an exponential fitting function: $T=a\exp(bN)$, with the parameters $a\simeq2.6\times10^{-7}$, $b\simeq1.08$ obtained from a best fit analysis in the large size regime $N\ge 14$. These results are obtained exploiting one core of Hardware1. The bond  dimension is scaled to reach the accuracy $R^2\ge 0.99$ for each $N$.}
	\label{mps_time}
\end{figure}

Given the above timings based on state-vector algorithms, tensor-network methods offer a promising alternative. In particular, for circuits that do not generate largely entangled states, the bond dimension needed to eliminate any bias is known to be small.
Shallow circuits acting on product input states are not expected to create strongly entangled states. It is thus worth inspecting whether the circuit depths addressed here are amenable to tensor-network simulations.
For this purpose, we determine the bond dimension required to predict the output expectation value $m_z$ with an accuracy corresponding to $R^2\ge 0.99$, as a function of qubit number and circuit depth.
The results are shown in Fig~\ref{MPS_var1024_mean4}, addressing a circuit featuring only intra-layer angle fluctuations, namely, the circuit configuration for which supervised deep-learning is efficient.
One notices that already for depths $P\gtrsim 30$ the bond dimension approaches the worst-case scenario, namely, the exponential scaling $\chi \propto 2^{N/2}$.
The small deviations from this scaling can be attributed to the allowed precision tolerance and to the finite circuit depth.
The exponential scaling of the bond dimension is expected to lead to an equivalent scaling of the total computation time of the tensor-network method. This is indeed demonstrated by the analysis reported in Fig.~\ref{mps_time},  which is based on Qiskit matrix-product-state routine executed on Hardware1.
Extrapolating the exponential trend observed for $N\gtrsim 14$ leads, again, to unfeasible timings for $N\simeq 60$ qubits.

All the timings reported above corroborate the findings discussed in Section~\ref{subsec3b}, which indicate that, for the circuits in configuration ii), the scalable CNNs can emulate quantum computers beyond the reach of the current implementations of the most popular classical simulation libraries. 

\section{Conclusions}\label{Sec5}
If and to what extent deep neural networks can emulate quantum circuits is an open question, with critical implications for the further development of quantum computing.
In fact, it was recently proposed that autoregressive networks, e.g., large language models, might be able to learn the complex correlations occurring in bitstrings measured from the output state of universal quantum computers~\cite{melko2024language}. Some numerical evidence  of this type of unsupervised training, which uses datasets including only measured bistrings, was indeed reported, but employing quantum-annealing devices tailored to sample low-energy configurations of spin glasses~\cite{10.21468/SciPostPhys.15.1.018}.
In the context of universal gate-based devices, deep neural networks were trained via supervised learning -- hence, using datasets featuring both circuit descriptors and the corresponding target values -- to predict some output expectation values~\cite{Cantori_2023,mohseni2023}. However, the limitations of this supervised approach are still unclear, and it is the purpose of this Article to quantify them in some relevant cases. In fact, by focusing on 
parametrized circuits familiar from applications of variational quantum algorithms, we identified a regime where supervised learning entails an exponential cost, as well as a successful regime where scalable networks outperform the most common general-purpose classical simulation algorithms.

In detail, the circuits we addressed feature layers of single-qubits rotations with random angles, alternated with layers of CNOT gates acting on all pairs of adjacent qubits on a chain. 
The variance of the (random) angle fluctuations turns out to be the critical factor that determines the performance of supervised learning.
Indeed, the required amount of training data scales exponentially with the inter-layer variance; this allows us identifying the regime where supervised-trained neural networks fail to efficiently emulate quantum circuits. It is worth mentioning that the heuristic circuits commonly used in the variational quantum eigensolver do indeed feature (also) inter-layer angle variations. Notice also that different neural networks, namely the multi-layer perceptron, convolutional networks, and long-short term memory networks, reach comparable performances, indicating that the amount of training data is the critical resource that determines the prediction accuracy.
On the other hand, circuits featuring only intra-layer angle fluctuations are accurately emulated at a feasible computational cost. In fact, thanks to the scalable architecture, the trained networks scale up to larger qubit numbers and circuit depths than those used in the training phase. 
Interestingly, the extrapolation accuracy rapidly improves with the size of the training circuits. In fact, by comparing predictions from trainings performed on different qubit numbers, we obtained clear indications that these extrapolations maintain an accuracy level of $R^2 \ge 0.99$ at least up to $N_{\mathrm{test}} \simeq 60$, and possibly even further. Notice that, as shown by the computational-cost analysis reported in Section~\ref{Sec4}, exactly simulating these circuit sizes is currently out of reach for two of the most popular and efficient state-vector simulation software. In fact, in the regime $N \gtrsim 20$, the computational cost displays the expected asymptotic exponential scaling with $N$. On the other hand, approximate simulations based on tensor-network method are also out of game, since for the circuit depths we consider the required bond dimension turns out to scale close to exponentially.

It is plausible that, in the near or midterm future, further algorithmic and/or hardware developments, in particular adopting some more specialized technique, might reach sizes comparable to the one addressed in this Article.  
For example, it was recently shown that, adopting the belief-propagation approximation for tensor contraction~\cite{PRXQuantum.5.010308}, a tensor network method is able to accurately simulate the 127 qubits of IBM's Eagle Kicked Ising experiment~\cite{ibm}.
To favour future comparisons with such novel simulation tools, as well as with actual quantum devices, we provide in the repository of Ref.~\cite{cantori_2024_10610695} dataset suitable for benchmarking purposes. In detail, these datasets include sets of circuits descriptors associated to the corresponding output expectation value $m_z$, as predicted by one of our scalable neural networks, reaching the circuit size $N=60$.

A full understanding of what makes a quantum circuit intractable for supervised learning will require further investigations. It is worth mentioning here that probabilistic deep-learning simulation algorithms have also been proposed~\cite{PhysRevA.104.032610}, but their reach is also still not well understood. The studies on supervised training are important also to set target goals to claim quantum computational advantage and to further improve error-mitigation schemes based on deep learning. In principle, supervised training could be performed also using a combination of exact expectation values from classical simulations of small circuits, and noisy outputs of (classically intractable) large-scale quantum devices. We leave these endeavours to future investigations.

\bmhead{Acknowledgements}
We are thankful to Luca Brodoloni, Emanuele Costa, Andrea Mari, Giuseppe Scriva, and David Vitali for fruitful discussions.
We acknowledge support from the Italian Ministry of University and Research under the PRIN2022 project ``Hybrid algorithms for quantum simulators'' -- 2022H77XB7, and partial support under the PRIN-PNRR 2022 MUR project "UEFA" -- P2022NMBAJ.
Support from the PNRR MUR project PE0000023-NQSTI is also acknowledged.
S.P. acknowledges support from the CINECA awards IsCa6\_NEMCAQS and IsCb2\_NEMCASRA, for the availability of high-performance computing resources and support.
We also acknowledge the EuroHPC Joint Undertaking for awarding this project access to the EuroHPC supercomputer LUMI, hosted by CSC (Finland) and the LUMI consortium through a EuroHPC Regular Access call.

\section*{Declarations}

\begin{itemize}
\item Some benchmark data are made available through the repository at Ref.~\cite{cantori_2024_10610695}. All other data and codes are available from the authors upon reasonable request.
\end{itemize}

\begin{appendices}

\section{Machine learning models}\label{appendix}
\subsection{Multilayer perceptron}
The multilayer perceptron (MLP) adopted in this study is composed of five layers. The first four layers include 512 neurons each. These layers employ the rectified linear unit (in jargon, ReLU) activation functions to add non-linearity. The output layer consists of a single neuron with linear activation.

\subsection{Long short-term memory network}
\begin{figure}[]
	\centering
	\includegraphics[width=0.5\columnwidth]{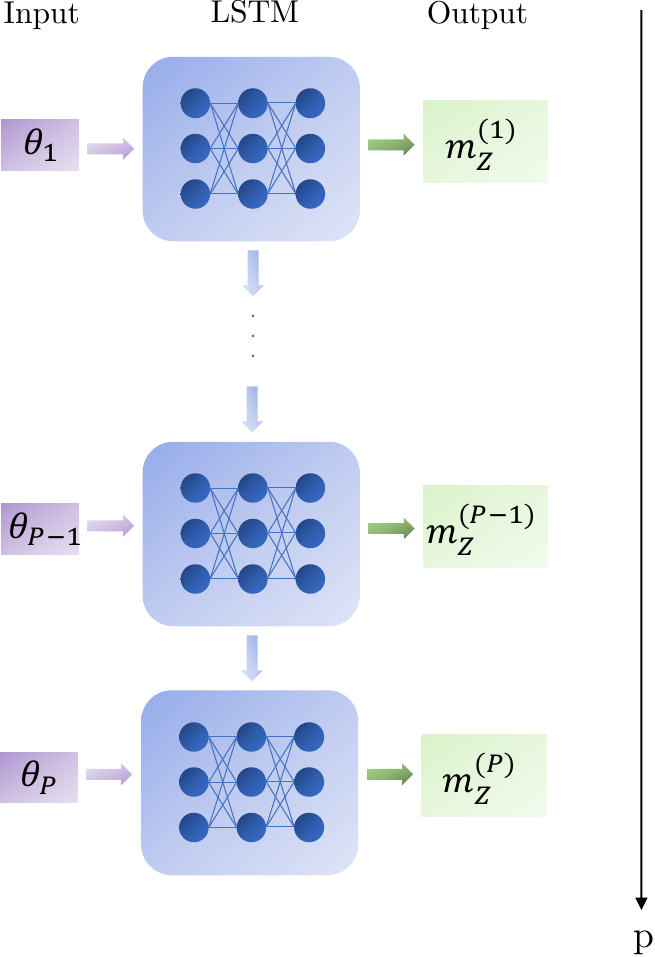}
	\caption{Schematic representation of the application of the LSTM network to the simulation of quantum circuit featuring $P$ layers. The rotation angle at layer $p=1,\dots,P$ is denoted with $\theta_p$, while $m_Z^{(p)}$ denote the magnetization per spin after the corresponding layer.}
	\label{LSTM}
\end{figure}

The long short-term memory (LSTM) network employed in this investigation is structured with three LSTM layers, each comprising 256 neurons. These layers are configured to produce an output for each time step, which is identified with the layer index $p$. The final layer is a dense layer that generates a value for each time step, containing the sequential information encoded by the preceding LSTM layers. In this case, the input to the $p$-th time-step is the angle describing the corresponding layer of gates, while the output is the global magnetization $m_z^{(p)}$ after the action of that layer of gates. This procedure is described in Fig.~\ref{LSTM}.
The basic architecture of an LSTM unit consists of three gates: the input gate ($i_t$), the forget gate ($f_t$), and the output gate ($o_t$), along with a memory cell ($c_t$) that stores information over time. The hidden state at time $t$ is denoted as $h_t$.
The input gate controls the flow of information that should be stored in the memory cell. It determines which parts of the incoming data should be remembered and updated in the cell state. The input gate is activated by the sigmoid function, which provides values between 0 and 1:
\begin{equation}
i_t = \sigma(W_{i}x_t + U_{i}h_{t-1} + b_{i}) \, .
\end{equation}
Here, $W_{i}$ and $U_{i}$ are weight matrices, $b_{i}$ is a bias vector, $\sigma$ is the sigmoid activation function, $x_t$ is the input at time $t$, and $h_{t-1}$ is the hidden state from the previous time step.
The forget gate decides which information from the previous cell state should be discarded. It helps in removing irrelevant information from the memory cell. The forget gate layer is
\begin{equation}
f_t = \sigma(W_{f}x_t + U_{f}h_{t-1} + b_{f}) \, ,
\end{equation}
where $W_{f}$ and $U_{f}$ are the weight matrices, while $b_{f}$ is the bias vector.
The output gate determines what information from the current memory cell state should be the output. It controls the flow of information from the memory cell to the hidden state. The output gate layer can be written as
\begin{equation}
o_t = \sigma(W_{o}x_t + U_{o}h_{t-1} + b_{o}) \, ,
\end{equation}
where $W_{o}$ and $U_{o}$ are the weight matrices and $b_{o}$ is the bias vector.
The memory cell is updated by combining the information from the input gate and the cell input activation ($g_t$), and forgetting the irrelevant information determined by the forget gate. The cell input activation vector and the state of the cell are described by the following equations:
\begin{align}
g_t &= \tanh(W_{g}x_t + U_{g}h_{t-1} + b_{g}) \, , \\
c_t &= f_t \odot c_{t-1} + i_t \odot g_t \, .
\end{align}
Here, $\tanh$ is the hyperbolic tangent activation function, $W_{g}$ and $U_{g}$ are weight matrices, $b_{g}$ is the bias vector, $\odot$ denotes element-wise multiplication, and $c_{t-1}$ is the previous cell state.
The hidden state is updated by applying the output gate to the hyperbolic tangent of the current cell state:
\begin{equation}
h_t = o_t \odot \tanh(c_t) \, .
\end{equation}

\subsection{Convolutional neural networks with flatten layer}
We use this artificial neural network as a comparison with MLP and LSTM. It has four convolutional layers with 32, 64, 128 and 128 filters each. They employ the ReLU activation function and a kernel size of three, with zero padding. A flatten layer connects the convolutional part with the dense part. The latter is composed by four layers, each containing 256 neurons and the ReLU activation function. The output layer includes one neuron with linear activation function.

\subsection{Convolutional neural networks with global pooling layer}
The 1D Convolutional neural network (1D-CNN) with global pooling includes four convolutional layers. The initial one features 256 filters, employs the ReLU activation function and utilizes a kernel size of three with zero padding. This layer is designed to process input sequences of variable length, each containing a single channel.
Successive convolutional layers maintain the same configuration.
Following the convolutional layers, a global average pooling layer is incorporated to capture global information across the data and to maintain the scalability property of the convolutional architecture.
The network ends with four dense layers, each containing 512 neurons and activated by the ReLU function.
The last dense layer consists of a single neuron,
      serving as the output layer for the network. 
The 2D convolutional neural network (2D-CNN) has an equivalent architecture but with 128 filters for each convolutional layer.

\end{appendices}

\bibliography{mybibliography}

\begin{thebibliography}{10}
\providecommand{\url}[1]{{#1}}
\providecommand{\urlprefix}{URL }
\providecommand{\doi}[1]{\url{https://doi.org/#1}}
\bibcommenthead

\bibitem{RevModPhys.91.045002}
G.~Carleo, I.~Cirac, K.~Cranmer, L.~Daudet, M.~Schuld, N.~Tishby, L.~Vogt-Maranto, L.~Zdeborov\'a, Machine learning and the physical sciences.
\newblock Rev. Mod. Phys. \textbf{91}, 045002 (2019).
\newblock \doi{10.1103/RevModPhys.91.045002}.
\newblock \urlprefix\url{https://link.aps.org/doi/10.1103/RevModPhys.91.045002}

\bibitem{doi:10.1080/23746149.2020.1797528}
J.~Carrasquilla, Machine learning for quantum matter.
\newblock Adv. Phys.: X \textbf{5}(1), 1797528 (2020).
\newblock \doi{10.1080/23746149.2020.1797528}.
\newblock \urlprefix\url{https://doi.org/10.1080/23746149.2020.1797528}.
\newblock {\href{https://arxiv.org/abs/https://doi.org/10.1080/23746149.2020.1797528}{{https://doi.org/10.1080/23746149.2020.1797528}}}

\bibitem{schutt2020machine}
K.T. Sch{\"u}tt, S.~Chmiela, O.A. Von~Lilienfeld, A.~Tkatchenko, K.~Tsuda, K.R. M{\"u}ller, Machine learning meets quantum physics.
\newblock Lect. Notes Phys.  (2020).
\newblock \doi{https://doi.org/10.1007/978-3-030-40245-7}

\bibitem{Kulik_2022}
H.J. Kulik, T.~Hammerschmidt, J.~Schmidt, S.~Botti, M.A.L. Marques, M.~Boley, M.~Scheffler, M.~Todorović, P.~Rinke, C.~Oses, A.~Smolyanyuk, S.~Curtarolo, A.~Tkatchenko, A.P. Bartók, S.~Manzhos, M.~Ihara, T.~Carrington, J.~Behler, O.~Isayev, M.~Veit, A.~Grisafi, J.~Nigam, M.~Ceriotti, K.T. Schütt, J.~Westermayr, M.~Gastegger, R.J. Maurer, B.~Kalita, K.~Burke, R.~Nagai, R.~Akashi, O.~Sugino, J.~Hermann, F.~Noé, S.~Pilati, C.~Draxl, M.~Kuban, S.~Rigamonti, M.~Scheidgen, M.~Esters, D.~Hicks, C.~Toher, P.V. Balachandran, I.~Tamblyn, S.~Whitelam, C.~Bellinger, L.M. Ghiringhelli, Roadmap on machine learning in electronic structure.
\newblock Electron. Struct. \textbf{4}(2), 023004 (2022).
\newblock \doi{10.1088/2516-1075/ac572f}.
\newblock \urlprefix\url{https://dx.doi.org/10.1088/2516-1075/ac572f}

\bibitem{doi:10.1126/science.aag2302}
G.~Carleo, M.~Troyer, Solving the quantum many-body problem with artificial neural networks.
\newblock Science \textbf{355}(6325), 602--606 (2017).
\newblock \doi{10.1126/science.aag2302}.
\newblock \urlprefix\url{https://www.science.org/doi/abs/10.1126/science.aag2302}

\bibitem{PRXQuantum.2.040201}
J.~Carrasquilla, G.~Torlai, How to use neural networks to investigate quantum many-body physics.
\newblock PRX Quantum \textbf{2}, 040201 (2021).
\newblock \doi{10.1103/PRXQuantum.2.040201}.
\newblock \urlprefix\url{https://link.aps.org/doi/10.1103/PRXQuantum.2.040201}

\bibitem{PhysRevLett.108.253002}
J.C. Snyder, M.~Rupp, K.~Hansen, K.R. M\"uller, K.~Burke, Finding density functionals with machine learning.
\newblock Phys. Rev. Lett. \textbf{108}, 253002 (2012).
\newblock \doi{10.1103/PhysRevLett.108.253002}.
\newblock \urlprefix\url{https://link.aps.org/doi/10.1103/PhysRevLett.108.253002}

\bibitem{https://doi.org/10.1002/qua.25040}
L.~Li, J.C. Snyder, I.M. Pelaschier, J.~Huang, U.N. Niranjan, P.~Duncan, M.~Rupp, K.R. Müller, K.~Burke, Understanding machine-learned density functionals.
\newblock Int. J. Quantum Chem. \textbf{116}(11), 819--833 (2016).
\newblock \doi{https://doi.org/10.1002/qua.25040}.
\newblock \urlprefix\url{https://onlinelibrary.wiley.com/doi/abs/10.1002/qua.25040}

\bibitem{Brockherde2017}
F.~Brockherde, L.~Vogt, L.~Li, M.E. Tuckerman, K.~Burke, K.R. M{\"u}ller, Bypassing the {Kohn-Sham} equations with machine learning.
\newblock Nat. Commun. \textbf{8}(1), 872 (2017).
\newblock \doi{10.1038/s41467-017-00839-3}.
\newblock \urlprefix\url{https://doi.org/10.1038/s41467-017-00839-3}

\bibitem{PhysRevE.106.045309}
E.~Costa, G.~Scriva, R.~Fazio, S.~Pilati, Deep-learning density functionals for gradient descent optimization.
\newblock Phys. Rev. E \textbf{106}, 045309 (2022).
\newblock \doi{10.1103/PhysRevE.106.045309}.
\newblock \urlprefix\url{https://link.aps.org/doi/10.1103/PhysRevE.106.045309}

\bibitem{PhysRevLett.98.146401}
J.~Behler, M.~Parrinello, Generalized neural-network representation of high-dimensional potential-energy surfaces.
\newblock Phys. Rev. Lett. \textbf{98}, 146401 (2007).
\newblock \doi{10.1103/PhysRevLett.98.146401}.
\newblock \urlprefix\url{https://link.aps.org/doi/10.1103/PhysRevLett.98.146401}

\bibitem{10.1063/1.4966192}
J.~Behler, {Perspective: Machine learning potentials for atomistic simulations}.
\newblock J. Chem. Phys. \textbf{145}(17), 170901 (2016).
\newblock \doi{10.1063/1.4966192}.
\newblock \urlprefix\url{https://doi.org/10.1063/1.4966192}

\bibitem{torlai2018neural}
G.~Torlai, G.~Mazzola, J.~Carrasquilla, M.~Troyer, R.~Melko, G.~Carleo, Neural-network quantum state tomography.
\newblock Nat. Phys. \textbf{14}(5), 447--450 (2018).
\newblock \doi{https://doi.org/10.1038/s41567-018-0048-5}

\bibitem{melko2024language}
R.G. Melko, J.~Carrasquilla, Language models for quantum simulation.
\newblock Nat. Comput. Sci. pp. 1--8 (2024).
\newblock \doi{https://doi.org/10.1038/s43588-023-00578-0}

\bibitem{Cantori_2023}
S.~Cantori, D.~Vitali, S.~Pilati, Supervised learning of random quantum circuits via scalable neural networks.
\newblock Quantum Sci. Technol. \textbf{8}(2), 025022 (2023).
\newblock \doi{10.1088/2058-9565/acc4e2}.
\newblock \urlprefix\url{https://dx.doi.org/10.1088/2058-9565/acc4e2}

\bibitem{C8SC04578J}
K.~Mills, K.~Ryczko, I.~Luchak, A.~Domurad, C.~Beeler, I.~Tamblyn, Extensive deep neural networks for transferring small scale learning to large scale systems.
\newblock Chem. Sci. \textbf{10}, 4129--4140 (2019).
\newblock \doi{10.1039/C8SC04578J}.
\newblock \urlprefix\url{http://dx.doi.org/10.1039/C8SC04578J}

\bibitem{saraceni}
N.~Saraceni, S.~Cantori, S.~Pilati, Scalable neural networks for the efficient learning of disordered quantum systems.
\newblock Phys. Rev. E \textbf{102}, 033301 (2020).
\newblock \doi{10.1103/PhysRevE.102.033301}.
\newblock \urlprefix\url{https://link.aps.org/doi/10.1103/PhysRevE.102.033301}

\bibitem{https://doi.org/10.1002/syst.201900052}
H.~Jung, S.~Stocker, C.~Kunkel, H.~Oberhofer, B.~Han, K.~Reuter, J.T. Margraf, Size-extensive molecular machine learning with global representations.
\newblock ChemSystemsChem \textbf{2}(4), e1900052 (2020).
\newblock \doi{https://doi.org/10.1002/syst.201900052}.
\newblock \urlprefix\url{https://chemistry-europe.onlinelibrary.wiley.com/doi/abs/10.1002/syst.201900052}

\bibitem{10.21468/SciPostPhys.10.3.073}
P.~Mujal, Àlex Martínez~Miguel, A.~Polls, B.~Juliá-Díaz, S.~Pilati, {Supervised learning of few dirty bosons with variable particle number}.
\newblock SciPost Phys. \textbf{10}, 073 (2021).
\newblock \doi{10.21468/SciPostPhys.10.3.073}.
\newblock \urlprefix\url{https://scipost.org/10.21468/SciPostPhys.10.3.073}

\bibitem{PhysRevB.108.125113}
E.~Costa, R.~Fazio, S.~Pilati, Deep learning nonlocal and scalable energy functionals for quantum ising models.
\newblock Phys. Rev. B \textbf{108}, 125113 (2023).
\newblock \doi{10.1103/PhysRevB.108.125113}.
\newblock \urlprefix\url{https://link.aps.org/doi/10.1103/PhysRevB.108.125113}

\bibitem{blania2022deep}
A.~Blania, S.~Herbig, F.~Dechent, E.~van Nieuwenburg, F.~Marquardt, Deep learning of spatial densities in inhomogeneous correlated quantum systems  (2022).
\newblock {\href{https://arxiv.org/abs/2211.09050}{{arXiv:2211.09050}}} {[quant-ph]}

\bibitem{PhysRevE.108.065304}
Z.~Tian, S.~Zhang, G.W. Chern, Machine learning for structure-property mapping of ising models: Scalability and limitations.
\newblock Phys. Rev. E \textbf{108}, 065304 (2023).
\newblock \doi{10.1103/PhysRevE.108.065304}.
\newblock \urlprefix\url{https://link.aps.org/doi/10.1103/PhysRevE.108.065304}

\bibitem{mohseni2023}
N.~Mohseni, J.~Shi, T.~Byrnes, M.~Hartmann, Deep learning of many-body observables and quantum information scrambling  (2023).
\newblock {\href{https://arxiv.org/abs/2302.04621}{{arXiv:2302.04621}}} {[quant-ph]}

\bibitem{Baireuther2018machinelearning}
P.~Baireuther, T.E. O'Brien, B.~Tarasinski, C.W.J. Beenakker, Machine-learning-assisted correction of correlated qubit errors in a topological code.
\newblock {Quantum} \textbf{2}, 48 (2018).
\newblock \doi{10.22331/q-2018-01-29-48}.
\newblock \urlprefix\url{https://doi.org/10.22331/q-2018-01-29-48}

\bibitem{Chamberland_2018}
C.~Chamberland, P.~Ronagh, Deep neural decoders for near term fault-tolerant experiments.
\newblock Quantum Sci. Technol. \textbf{3}(4), 044002 (2018).
\newblock \doi{10.1088/2058-9565/aad1f7}.
\newblock \urlprefix\url{https://dx.doi.org/10.1088/2058-9565/aad1f7}

\bibitem{Baireuther_2019}
P.~Baireuther, M.D. Caio, B.~Criger, C.W.J. Beenakker, T.E. O’Brien, Neural network decoder for topological color codes with circuit level noise.
\newblock New J. Phys. \textbf{21}(1), 013003 (2019).
\newblock \doi{10.1088/1367-2630/aaf29e}.
\newblock \urlprefix\url{https://dx.doi.org/10.1088/1367-2630/aaf29e}

\bibitem{zlokapa2020deep}
A.~Zlokapa, A.~Gheorghiu, A deep learning model for noise prediction on near-term quantum devices  (2020).
\newblock {\href{https://arxiv.org/abs/2005.10811}{{arXiv:2005.10811}}} {[quant-ph]}

\bibitem{8880492}
S.~Varsamopoulos, K.~Bertels, C.G. Almudever, Comparing neural network based decoders for the surface code.
\newblock IEEE Trans. Comput. \textbf{69}(2), 300--311 (2020).
\newblock \doi{10.1109/TC.2019.2948612}

\bibitem{ml_qem}
H.~Liao, D.S. Wang, I.~Sitdikov, C.~Salcedo, A.~Seif, Z.K. Minev, {Machine learning for practical quantum error mitigation}  (2023).
\newblock {\href{https://arxiv.org/abs/2309.17368}{{arXiv:2309.17368}}} {[quant-ph]}

\bibitem{doi:10.1021/acs.jctc.1c00091}
F.~Benfenati, G.~Mazzola, C.~Capecci, P.K. Barkoutsos, P.J. Ollitrault, I.~Tavernelli, L.~Guidoni, Improved accuracy on noisy devices by nonunitary variational quantum eigensolver for chemistry applications.
\newblock J. Chem. Theory Comput. \textbf{17}(7), 3946--3954 (2021).
\newblock \doi{10.1021/acs.jctc.1c00091}.
\newblock \urlprefix\url{https://doi.org/10.1021/acs.jctc.1c00091}

\bibitem{PhysRevA.102.062612}
Y.S. Yordanov, D.R.M. Arvidsson-Shukur, C.H.W. Barnes, Efficient quantum circuits for quantum computational chemistry.
\newblock Phys. Rev. A \textbf{102}, 062612 (2020).
\newblock \doi{10.1103/PhysRevA.102.062612}.
\newblock \urlprefix\url{https://link.aps.org/doi/10.1103/PhysRevA.102.062612}

\bibitem{Barkoutsos2020improving}
P.K. Barkoutsos, G.~Nannicini, A.~Robert, I.~Tavernelli, S.~Woerner, Improving {V}ariational {Q}uantum {O}ptimization using {CV}a{R}.
\newblock {Quantum} \textbf{4}, 256 (2020).
\newblock \doi{10.22331/q-2020-04-20-256}.
\newblock \urlprefix\url{https://doi.org/10.22331/q-2020-04-20-256}

\bibitem{shen2023prepare}
Y.~Shen, Prepare ansatz for vqe with diffusion model  (2023).
\newblock {\href{https://arxiv.org/abs/2310.02511}{{arXiv:2310.02511}}} {[quant-ph]}

\bibitem{scriva2023challenges}
G.~Scriva, N.~Astrakhantsev, S.~Pilati, G.~Mazzola, Challenges of variational quantum optimization with measurement shot noise  (2023).
\newblock {\href{https://arxiv.org/abs/2308.00044}{{arXiv:2308.00044}}} {[quant-ph]}

\bibitem{cantori_2024_10610695}
S.~Cantori, S.~Pilati.
\newblock {Challenges and opportunities in the supervised learning of quantum circuits} (2024).
\newblock \doi{10.5281/zenodo.10610695}

\bibitem{Zhang_2021}
S.X. Zhang, C.Y. Hsieh, S.~Zhang, H.~Yao, Neural predictor based quantum architecture search.
\newblock Mach. Learn.: Sci. Technol. \textbf{2}(4), 045027 (2021).
\newblock \doi{10.1088/2632-2153/ac28dd}.
\newblock \urlprefix\url{https://dx.doi.org/10.1088/2632-2153/ac28dd}

\bibitem{hastie2009elements}
T.~Hastie, R.~Tibshirani, J.H. Friedman, J.H. Friedman, \emph{The elements of statistical learning: data mining, inference, and prediction}, vol.~2 (Springer, New York, NY, 2009).
\newblock \doi{https://doi.org/10.1007/978-0-387-21606-5}

\bibitem{10.1162/neco.1997.9.8.1735}
S.~Hochreiter, J.~Schmidhuber, {Long Short-Term Memory}.
\newblock Neural Comput. \textbf{9}(8), 1735--1780 (1997).
\newblock \doi{10.1162/neco.1997.9.8.1735}

\bibitem{lstm}
N.~Mohseni, T.~Fösel, L.~Guo, C.~Navarrete-Benlloch, F.~Marquardt, Deep learning of quantum many-body dynamics via random driving.
\newblock Quantum \textbf{6}, 714 (2022).
\newblock \doi{10.22331/q-2022-05-17-714}.
\newblock \urlprefix\url{https://doi.org/10.22331%2Fq-2022-05-17-714}

\bibitem{global_pooling2}
N.~Mohseni, C.~Navarrete-Benlloch, T.~Byrnes, F.~Marquardt, Deep recurrent networks predicting the gap evolution in adiabatic quantum computing.
\newblock Quantum \textbf{7}, 1039 (2023).
\newblock \doi{10.22331/q-2023-06-12-1039}.
\newblock \urlprefix\url{https://doi.org/10.22331%2Fq-2023-06-12-1039}

\bibitem{venkatesan2017convolutional}
R.~Venkatesan, B.~Li, \emph{Convolutional neural networks in visual computing: a concise guide} (CRC Press, Boca Raton, 2017).
\newblock \doi{https://doi.org/10.4324/9781315154282}

\bibitem{qiskit}
{Qiskit contributors}.
\newblock Qiskit: An open-source framework for quantum computing (2023).
\newblock \doi{10.5281/zenodo.2573505}

\bibitem{qsimcirq}
Q.A. team, collaborators.
\newblock qsim (2020).
\newblock \doi{10.5281/zenodo.4023103}.
\newblock \urlprefix\url{https://doi.org/10.5281/zenodo.4023103}

\bibitem{adam}
D.P. Kingma, J.~Ba, Adam: A method for stochastic optimization  (2017).
\newblock {\href{https://arxiv.org/abs/1412.6980}{{arXiv:1412.6980}}} {[cs.LG]}

\bibitem{ibm}
Y.~Kim, A.~Eddins, S.~Anand, K.X. Wei, E.~van~den Berg, S.~Rosenblatt, H.~Nayfeh, Y.~Wu, M.~Zaletel, K.~Temme, A.~Kandala, Evidence for the utility of quantum computing before fault tolerance.
\newblock Nature \textbf{618}(7965), 500--505 (2023).
\newblock \doi{https://doi.org/10.1038/s41586-023-06096-3}.
\newblock \urlprefix\url{https://doi.org/10.1038/s41586-023-06096-3}

\bibitem{jamadagni2024benchmarking}
A.~Jamadagni, A.M. Läuchli, C.~Hempel, Benchmarking quantum computer simulation software packages  (2024).
\newblock {\href{https://arxiv.org/abs/2401.09076}{{arXiv:2401.09076}}} {[quant-ph]}

\bibitem{10.21468/SciPostPhys.15.1.018}
G.~Scriva, E.~Costa, B.~McNaughton, S.~Pilati, {Accelerating equilibrium spin-glass simulations using quantum annealers via generative deep learning}.
\newblock SciPost Phys. \textbf{15}, 018 (2023).
\newblock \doi{10.21468/SciPostPhys.15.1.018}.
\newblock \urlprefix\url{https://scipost.org/10.21468/SciPostPhys.15.1.018}

\bibitem{PRXQuantum.5.010308}
J.~Tindall, M.~Fishman, E.M. Stoudenmire, D.~Sels, Efficient tensor network simulation of ibm's eagle kicked ising experiment.
\newblock PRX Quantum \textbf{5}, 010308 (2024).
\newblock \doi{10.1103/PRXQuantum.5.010308}.
\newblock \urlprefix\url{https://link.aps.org/doi/10.1103/PRXQuantum.5.010308}

\bibitem{PhysRevA.104.032610}
J.~Carrasquilla, D.~Luo, F.~P\'erez, A.~Milsted, B.K. Clark, M.~Volkovs, L.~Aolita, Probabilistic simulation of quantum circuits using a deep-learning architecture.
\newblock Phys. Rev. A \textbf{104}, 032610 (2021).
\newblock \doi{10.1103/PhysRevA.104.032610}.
\newblock \urlprefix\url{https://link.aps.org/doi/10.1103/PhysRevA.104.032610}

\end{thebibliography}

\end{document}